\documentstyle[11pt,newpasp,twoside,epsfig]{article}
\begin{document}

\title{Stream eclipse mapping with 'fire-flies'}

\author{C. M. Bridge$^1$, Pasi Hakala$^2$, Mark Cropper$^1$, Gavin Ramsay$^1$}
\affil{$^1$Mullard Space Science Laboratory, University College London, Holmbury
St. Mary, Dorking, Surrey, RH5 6NT}
\affil{$^2$Observatory, University of Helsinki, Finland}

\begin{abstract}
We apply a new method of eclipse mapping to the light curves of eclipsing
polars. The technique aims to locate the bright emission associated with the
accretion stream, using a technique that makes the fewest prior assumptions
about the location of the accretion stream material. We have obtained data of
EP~Dra and HU~Aqr with the S-Cam~2 superconducting tunnel junction camera using
the William Herschel Telescope. The location of emission regions in both
systems show that previously assumed trajectories are consistent with those
found using our technique. Most of the emission is located in a region where we
expect material to be confined to magnetic field lines, particularly for HU
Aqr, while there appears to be less emission from where we conventionally
expect material to follow a ballistic trajectory from the $L_1$ point.
\end{abstract}

\section{Introduction}

Eclipse mapping is an inversion technique that can be used to reconstruct the
distribution of bright material in the accretion streams of eclipsing
polars. The eclipsing nature of these systems means that for a short duration
the stream is the dominant contributor to the observed brightness, the white
dwarf and accretion region being hidden from view by the secondary. The
technique of eclipse mapping was first applied to accretion disks in the early
1980s, and the method was later applied to the bright eclipsing polar HU~Aqr by
Hakala (1995). Because it was found that the accretion stream contributed
around half the optical emission in HU~Aqr, this system was ideal for the
application of eclipse mapping to determine the distribution of this bright
stream material in the binary system.

Hakala (1995) developed this method assuming a one-dimensional stream confined
to the orbital plane. This was represented as an arc connecting the inner
Lagrangian point ($L_1$) to the white dwarf. Developments to this method
included a two part trajectory, and models using this have been successfully
applied to the eclipse light curves of the bright eclipsing polars HU~Aqr and
UZ~For (Bridge et al. 2002; Vrielmann \& Schwope 2001; Harrop-Allin, Potter \&
Cropper 2001; Kube, G\"ansicke \& Beuermann 2000; Harrop-Allin et
al. 1999b). This trajectory comprises a ballistic free-fall component from the
$L_1$-point (Lubow \& Shu 1975), followed at some distance from the white dwarf
by a magnetically confined component. The confined part of the trajectory
originates in the threading region from where the stream material is channeled
by the field lines to the white dwarf surface. The field lines are assumed to
have a dipolar geometry and be centred on the white dwarf. However, such a
clear division between two trajectories is likely to be an inadequate
assumption (e.g. Heerlein, Horne \& Schwope 1999) and the assumption of a
ballistic free-fall component may not be appropriate in all cases (e.g. the low
state of HU~Aqr; Harrop-Allin et al. 2001). A further development is a move to
three-dimensional streams (Vrielmann \& Schwope 2001; Kube et al. 2000), thus
attempting to include the effects of uneven heating of the stream and features
related to the eclipse by the stream of the white dwarf and accretion region.

All these previous model techniques have the common feature that the stream
trajectory is fixed prior to the application of the model. Hakala, Cropper \&
Ramsay (2002) introduced a technique which makes no assumptions about the
location of the stream material prior to the modeling process. In principle,
emission is allowed from anywhere within the Roche lobe of the white dwarf.
They have successfully applied this technique to synthetic data sets, but here
we apply it to observed light curves for the first time.

We model the `white' light curves of the eclipsing polars EP~Dra and HU~Aqr
(see Bridge et al. 2002, 2003 for a discussion of the light curves),
observations of which were obtained with the superconducting tunnel junction
(STJ) camera S-Cam 2 (see e.g. Perryman et al. 2001). The S-Cam 2 camera
records the location on the array, time of arrival and energy of each incident
photon. This allows the study of the energy dependence of short timescale
variations observed in polars.

\begin{figure}
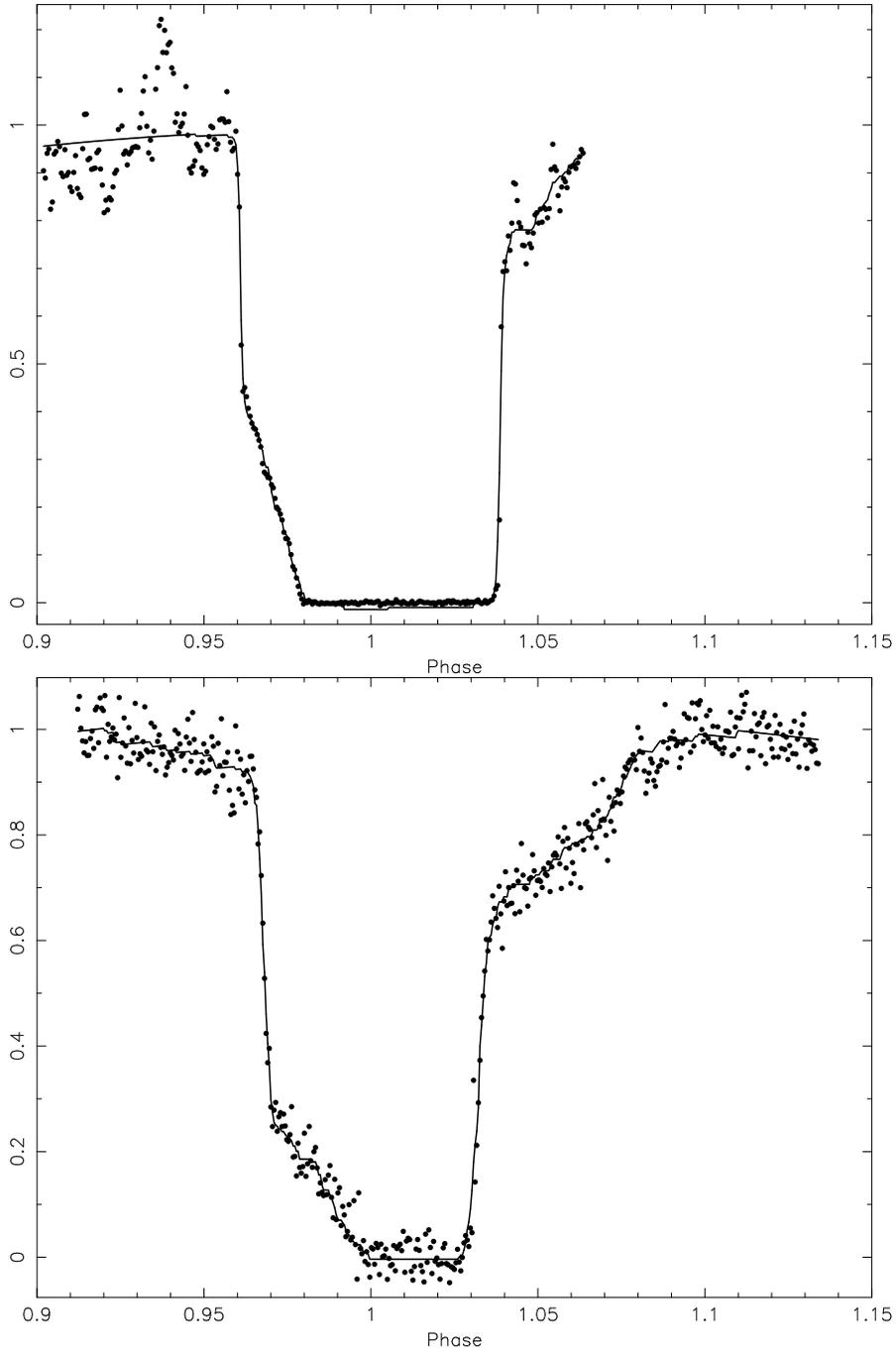

\centerline{\epsfig{file=./cmbridge_fig1a.ps,width=9.cm,angle=-90}}
\centerline{\epsfig{file=./cmbridge_fig1b.ps,width=9.cm,angle=-90}}
\caption{The model fits (solid line) to the observed light curves
(dots) for HU~Aqr (top) and EP~Dra (bottom). The light curves are
binned in 3~s intervals and normalised to 1 pre-eclipse.}
\label{fig:fig01}
\end{figure}

\section{Model technique}

We discuss briefly here the main points of the model (for a detailed discussion
see Hakala et al. 2002). The model consists of a number of `fire-flies', where
each fly is an individual bright emission point. A fly is created with an
initially random location within the Roche lobe of the white dwarf, but is
subsequently free to move during the evolution of the model. The fly has a
brightness ($F_{fly}$) which varies with the angle $\alpha$ between the
observer, the fly and the white dwarf. This is defined as
$F_{fly}(\alpha)=F_{0}+Acos(\alpha)$ where $F_{0}$ is the minimum brightness of
a fly and $A$ is the amplitude of angular dependence. The brightness of those
flies not eclipsed by the secondary at each phase are summed to form the model
light curve. Many swarms of such flies are created and for each a model light
curve can be generated.

The goodness of fit of an individual model light curve to the observed light
curve is then defined using a fitness function, $R$, which consists of a
$\chi^2$ term and a regularisation term: $R=\chi^2 + \lambda S_{reg}$. The best
fit solution is `bred' through a number of generations using a genetic
algorithm (GA; see e.g. Charbonneau 1995 for a review) that is regulated by a
self-organising map (SOM; Kohonen 1990). The regularisation places a number of
sections of a smooth curve through the fly distribution (see Hakala et
al. 2002), this helps to constrain the model where generally there are more
degrees of freedom than data points. The two end points of the curve are
defined as the $L_1$-point and the white dwarf surface, thus providing two
physically realistic constraints on the stream trajectory. The model then
prefers those swarms with minimal distance to the curve, resulting in smooth
tube-like fly distributions. Hakala et al. (2002) also demonstrated the use of
the technique with no regularisation term. Here there is no computational
pressure on the flies to create any form of tube and they are not constrained
to lie between the white dwarf and the $L_1$-point. In this case the model will
define those regions of the Roche lobe volume where emission can and cannot be.

\begin{figure}
\centerline{\epsfig{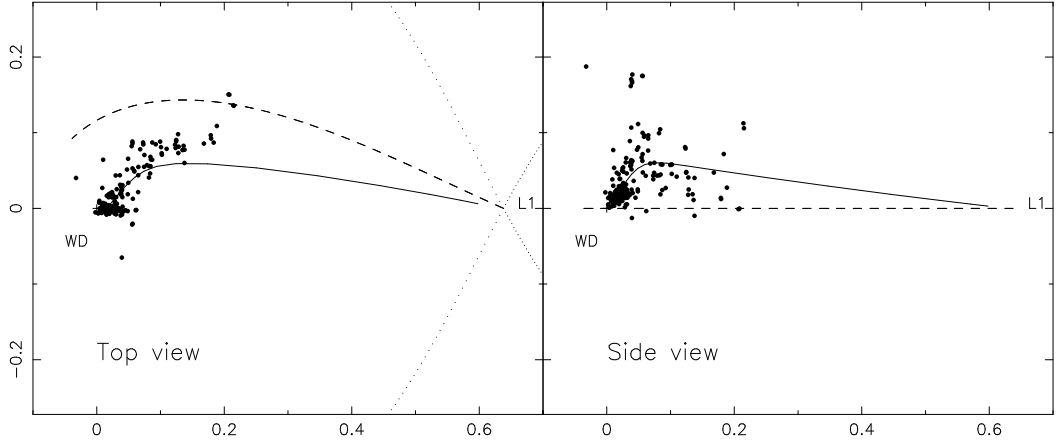}}
\caption{The fly distribution for HU~Aqr defining the emission region. The
left-hand panel shows the x-y plane looking down onto the orbital
plane, while the right-hand panel shows the view parallel to the
orbital plane. The dotted line illustrates the Roche lobe, the
dashed line a ballistic trajectory and the solid line the
regularisation curve.} 
\label{fig:fig02}
\end{figure}

\section{Results}

\begin{figure}
\centerline{\epsfig{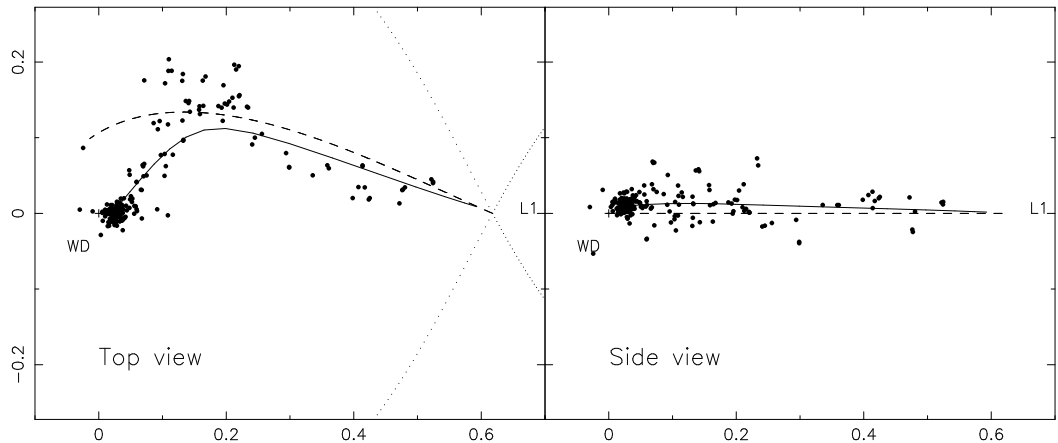}}
\caption{As for Figure~\ref{fig:fig02} but for EP~Dra.} 
\label{fig:fig03}
\end{figure}

The light curves were obtained in October 2000 at the William Herschel
Telescope, La Palma.  Preliminary results for fits to the `white' light curves
of EP~Dra and HU~Aqr are shown in Figure~\ref{fig:fig01}, the light curves are
binned into 3~s time intervals, with the model fits superimposed as a solid
line. We used 500 swarms each with 200 flies, and the models are run for 500
generations. This provides a sufficient number of swarms to produce a unique
solution and enough iterations to reach convergence. Figures~\ref{fig:fig02}
and \ref{fig:fig03} show the corresponding fly distributions.

The fitting process is necessarily a trade-off, mediated by the value of
$\lambda$, between fitting real gradient changes and avoiding statistical noise
features. The presence of noise in the data will have the effect of broadening
the fly distribution, but we can attempt to minimise this by adjusting the
$\lambda$ term to place more emphasis on the regularisation. However, while
increasing $\lambda$ has the effect of smoothing the distribution of flies into
a more tube-like structure, we must be careful not to place too much emphasis
on $\lambda$ as this will have the effect of smoothing out any real brightness
variations.

Figure~\ref{fig:fig02} shows that the emission points in HU~Aqr are
located mostly out of the orbital plane. The extent in volume of the emission
points, and location at different distances in the z-direction may indicate
that material is threaded along many different field lines, and may be the
cause of the lack of evidence for a bright threading region at any one place in
the orbital plane. In contrast to HU~Aqr, the emission points in EP~Dra are
located close to the orbital plane, possibly indicating accretion along field
lines that meet the white dwarf surface at low latitudes. The emission in
EP~Dra is also distributed over a larger volume than HU~Aqr, particularly the
region where we expect coupling to field lines to occur. To some extent this is
a product of a noisier light curve, but it could also indicate material located
in a large stagnation region, where it collects before coupling to the white
dwarf field lines.

\section{Discussion}

We have applied this more objective technique to real data in an attempt to
make the least number of assumptions about the location of bright material in
these eclipsing binary systems. We find that previous assumptions are largely
consistent with our results. For both EP~Dra and HU~Aqr most of the emission is
located in the region where we expect material to be confined to the magnetic
field lines of the white dwarf. In particular both models show an enhanced
concentration of emission points towards the white dwarf. This may be
indicative of where stream material is heated by X-ray emission from the
accretion region on the white dwarf. However, the contribution from the
accretion region, seen as the large amplitude change at ingress and egress, is
fixed for the duration of the model. Therefore, if the input value is too low
the model will assign flies to the white dwarf to boost the contribution at the
ingress and egress of the accretion region. The number of flies near to the
white dwarf may be indicative of this effect, rather than a real stream
brightness enhancement.

An important advantage of placing no constraint on the stream trajectory, is
the lack of sensitivity to input parameters. The HU~Aqr data presented here has
been modeled previously with a version of the technique of Harrop-Allin et
al. (1999a). This method was found to be acutely sensitive to the exact value
chosen for the distance to the threading region, and also the location of the
accretion region on the white dwarf surface, which fixes the geometry of the
magnetic field lines (Bridge et al. 2002). This meant that the application to
the data was restricted, and results had to be interpreted accordingly. As this
new technique does not rely on the choice of initial parameters for the stream
trajectory, we reduce these uncertainties.

The stream trajectory used previously by Bridge et al. (2002) to model the
HU~Aqr data is consistent with that found from this new technique. The results
of this modeling identified brightness enhancements towards the white dwarf and
in the threading region.  However with this new technique we do not see much
enhancement where we expect material to be threaded by the field lines. The
technique of Harrop-Allin et al. (1999a) was also applied to other observations
of HU~Aqr (Harrop-Allin et al. 1999b). They found that there was generally
enhanced brightness towards the white dwarf and in the threading region, and
for some cycles the threading region was significantly brighter. By comparison
to both these methods, Vrielmann \& Schwope (2000) did not find any enhancement
towards the white dwarf, but did identify a brighter threading region. However,
caution should be exercised as they used line emission only, while here we use
the total emission (line and continuum). The results from modeling different
light curves of HU~Aqr may indicate some evolution of the stream brightness
with time. However, the results of applying different models to the same data
indicate that the results may be dependent upon the model parameters, as found
for the technique of Harrop-Allin et al. (Bridge et al. 2002).

In HU~Aqr we see no evidence for emission from where we would expect material
to be following a ballistic trajectory. In EP~Dra however, there is some
indication that there is emission from along a ballistic trajectory (in
Figures~\ref{fig:fig02} and \ref{fig:fig03} the dashed curve represents the
ballistic trajectory, seen most clearly in the left-hand panels). The location
of material close to the orbital plane in EP~Dra causes accretion at low
latitude and is consistent with the suggested location of the accretion region
from the cyclotron models of Schwope \& Mengel (1997). Low latitude accretion
may be indicative of a more complicated magnetic field geometry, with a
quadrupole more likely to cause accretion at lower latitudes (Wu \&
Wickramasinghe 1993).

\section{Summary and further work}

While the previous techniques all assumed a stream trajectory, here we use a
less constrained model that makes few {\it a priori} assumptions about the
location of material. This shows that the trajectory is largely as expected,
with the model providing good fits to the light curves of HU~Aqr and
EP~Dra. The different results obtained from modeling the same data imply some
dependence upon the technique, although this new technique is less sensitive to
input parameters than that of Harrop-Allin et al. (1999a). A more detailed
investigation of the effects of the input parameters and the regularisation
curve is needed to fully understand any effects. We see most of the emission
from material close to the white dwarf in both systems, and it is unclear why
we see no emission from where we expect material to be following a ballistic
trajectory in HU~Aqr.

The colour information available from S-Cam 2 means that we can use the model
to determine the colour dependence of the stream emission. In particular this
will help explore the temperature variations along the stream by emphasising
which parts of the stream are brightest at bluer wavelengths, and therefore
hotter, and where the stream is redder and hence cooler. We can also apply the
model to the consecutive cycles of HU~Aqr and EP~Dra, to examine any evolution
of the stream structure and temperature.

\end{document}